# Analog Optical Computing Primitives in Silicon Photonics

Yunshan Jiang[1,*], Peter T.S. DeVore[1,2], Bahram Jalali[1]

[1]*Electrical Engineering Department, UCLA, Los Angeles, CA, USA*
[2]*Engineering Directorate, Lawrence Livermore National Laboratory, Livermore, CA, USA*
*Corresponding author: yunshanjiang@gmail.com*

Optical computing accelerators may help alleviate bandwidth and power consumption bottlenecks in electronics. We show an approach to implementing logarithmic-type analog co-processors in silicon photonics and use it to perform the exponentiation operation. The function is realized by exploiting nonlinear-absorption-enhanced Raman amplification saturation in a silicon waveguide.

OCIS Codes: (200.4560) Optical data processing, (190.4390) Nonlinear optics, (130.0130) Integrated optics.
http://

With the proliferation of big data and the rapid increase in power dissipation of electronics, there is renewed interest in the use of optics for computing. In contrast to the optical computing efforts of the past [1,2], an all-optical computer may not be the most prudent goal. Instead, a hybrid approach where optical systems are selectively applied to alleviate bottlenecks and assist electronic processors is a more fruitful pursuit. The idea of optical co-processors is proposed as hardware accelerators to take part of the processing burden off of the electronic processors, as shown in Fig. 1(a) [3]. Composed of carefully designed photonics components, the optical co-processor performs a certain analog computational operation in real time on the input optical signal before it is acquired and digitized.

Among the analog-computing primitives, the logarithmic function is of importance and is one of the most challenging operations to perform in optics. As illustrated in Fig. 1 (b), an optical implementation of the exponentiation operation can be achieved by three sequential components: the logarithmic primitive, the scaling primitive, and the natural exponentiation primitive. Apart from the logarithmic primitive, the two remaining primitives can be emulated using commercially available optical systems. For example, scaling can be achieved with variable optical attenuators or four-wave mixing, while a Raman amplifier operating in the low depletion regime provides the natural exponentiation function with respect to the input intensity.

The lack of logarithmic dependence in conventional optical interactions renders the realization of a logarithm computation block formidable. Logarithmic filtering was demonstrated in literature using nonlinear photographic films [4] and hologram masks [5], but the cumbersome free space setup and the complicated processing have limited its range of applicability.

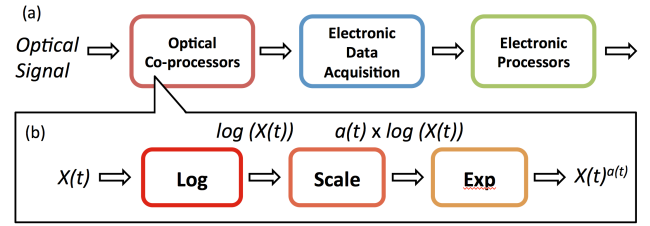

Fig. 1. (a) Optical co-processors that perform computational operations on optical input signal can be placed before the optical-to-electrical conversion to take part of the processing burden off of the electronic processors [3]. (b) As one of the building blocks of the optical co-processors, the exponentiation operation is composed of the Logarithmic Primitive (Log), the Scale Primitive (Scale) and the Exponential Primitive (Exp).

In this paper, we show an approach to approximate the optical input-output relationship as a logarithmic function in a silicon waveguide via numerical studies. Silicon naturally exhibits two-photon absorption (TPA) at telecommunication wavelengths. This nonlinear absorption, which limits the signal's output intensity and is normally deleterious [6], becomes a fortuitous natural candidate to approximate the logarithm. In the presence of a suitably wavelength-shifted pump source, stimulated Raman scattering amplifies the signal, and it has such a strong effect that it caused this indirect band-gap material to lase for the first time [7]. When the signal grows strong enough to deplete the pump source, the gain saturates, leading to a second method to achieve logarithm-like behavior. Non-degenerate TPA is also introduced to the system to shape the output curve. Using several effects simultaneously allows one to engineer their relative strengths, improving the dynamic range and lowering the required signal intensity.

For a quasi-continuous signal with wavelength below silicon band edge, two-photon absorption (TPA) and the induced free-carrier absorption (FCA) are the main sources of nonlinear loss in silicon waveguides [8]. The

evolution of optical intensity along the waveguide is described as [9]:

$$\frac{dI_s}{dz} = -\alpha I_s - \beta_{TPA}I_s^2 - \sigma \Delta N I_s \quad (1)$$

where $\alpha$ is the linear loss coefficient, $\beta_{TPA} = 5*10^{-12}$ m/W is the TPA coefficient, which is proportional to the imaginary part of third-order susceptibility, and $\sigma = 1.45*10^{-21}$ m$^2$ is the cross section of free carrier absorption. At steady state, the free carrier density $\Delta N$ is represented by

$$\Delta N = \frac{\tau \beta_{TPA}}{2h\nu_0}I_s^2 \quad (2)$$

where $\tau$ is the free carrier lifetime, and $h\nu_0$ is the photon energy.

The optical limiting phenomenon is observed at high input intensity as a result of the dominant nonlinear loss [10]. Between the linear region and the saturation region, there exists a sublinear curve that resembles a logarithmic function, as illustrated in Fig. 2. The logarithmic region is defined as the largest input intensity range whose output can be fit to a logarithmic function.

As a measurement for the fitting accuracy, two deviation calculation methods are employed. To evaluate the average accuracy of the computing primitive, the normalized root-mean-square error (NRMSE) should be no larger than 1% and is defined as

$$NRMSE = \frac{\sqrt{<(I_{out}-I_{fit})^2>}}{I_{max}-I_{min}} \quad (3)$$

To ensure the accuracy of each single input value, the maximum error should be no larger than 3.5% and is defined as

$$Max\ Error = max\left(\frac{|I_{out}-I_{fit}|}{I_{out}}\right) \quad (4)$$

An example of waveguide with length $Z = 2$ cm, lifetime $\tau = 1$ ns, and propagation loss $\alpha = 3$ dB/cm is shown in Fig. 2. The signal undergoes degenerate TPA and FCA and is fit to a logarithmic function $I_{fit}$ over the input intensity from 50 MW/cm$^2$ to 250 MW/cm$^2$, resulting in 7 dB dynamic range. It is noted that it requires very high input power to reach the logarithmic region. This results from the large ratio between the linear loss coefficient and the nonlinear loss coefficient: the nonlinear term only comes into effect when input intensity is above a certain region. A low propagation loss coefficient and a large free-carrier lifetime would reduce this ratio and shift the logarithmic region to lower input intensity. Unfortunately, a large free-carrier lifetime is not practical because it also reduces device's speed, while ultra-low linear absorption is limited by the fabrication technology. A practical computing primitive thus would require larger logarithmic range, lower power, and more flexibility.

Stimulated Raman Scattering offers optical gain in silicon without requiring phase matching [11]. The saturation of Raman amplification provides the opportunity to reach the logarithmic region with low input signal power. It also increases the dynamic range without significantly increasing setup complexity.

The Raman amplification in silicon along with nonlinear absorption can be modeled as [6]:

$$\begin{cases}\frac{dI_s}{dz} = (-\alpha + g_R I_R)I_s - \beta_{TPA}(I_s + 2I_R)I_s - \sigma \Delta N I_s \\ \frac{dI_R}{dz} = \left(-\alpha - \frac{\lambda_s}{\lambda_R}g_R I_s\right)I_R - \beta_{TPA}(I_R + 2I_s)I_R - \sigma \Delta N I_R \quad (5) \\ \Delta N = \frac{\tau_c \beta_{TPA}}{2h\nu_0}(I_s^2 + I_R^2 + 2I_sI_R)\end{cases}$$

where $g_R = 76$ cm/GW is Raman gain coefficient [6], and $I_R$ is the Raman pump intensity. Without loss of generality, the wavelength dependence of the linear loss coefficient $\alpha$, TPA coefficient $\beta_{TPA}$, and FCA coefficient $\sigma$ are ignored.

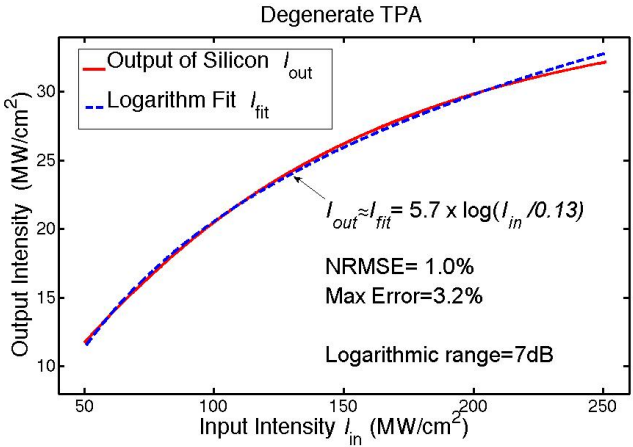

Fig. 2. Numerical demonstration of the logarithmic computing primitive in a silicon waveguide. The signal undergoes degenerate two-photon absorption (TPA) and free-carrier absorption (FCA). The output is fit to a logarithmic function over a 7 dB input range with a normalized-mean-square error is 1.0% and the maximum error is 3.2%.

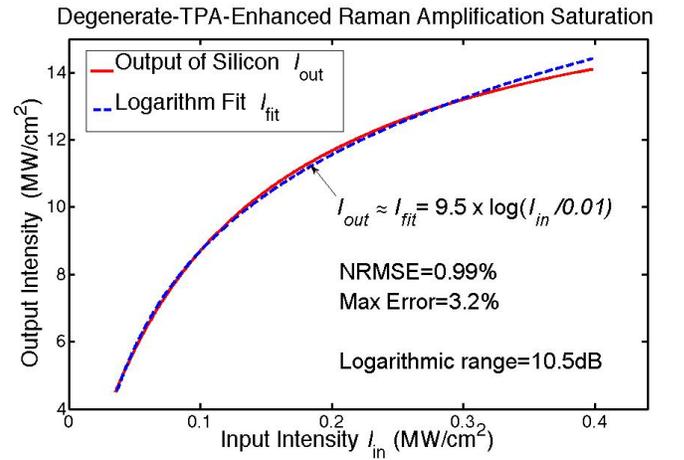

Fig. 3. Similar to Fig. 2, a simulation of the logarithmic computing primitive wherein Raman amplification along with concomitant non-degenerate TPA is added to increase the dynamic range and vastly reduce required signal intensity. The input Raman pump intensity is 91 MW/cm$^2$. The output is fit to a logarithmic function over a 10.5 dB input range with a normalized-mean-square error is 1.0% and the maximum error is 3.2%.

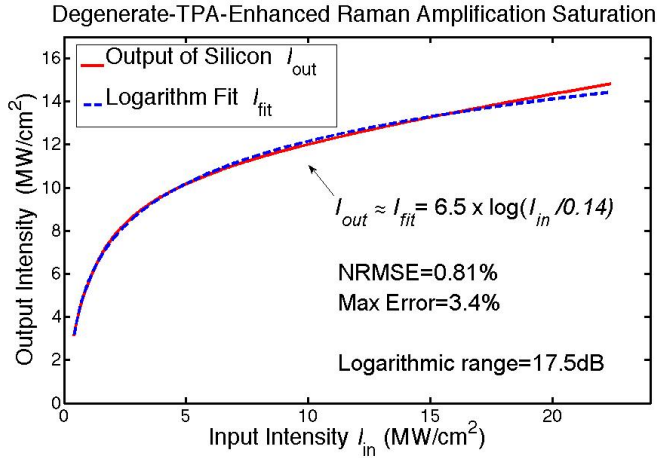

Fig. 4. Simulation of the logarithmic computing primitive under the same conditions as in Figure 3 (a Raman pump is added and undergoes non-degenerate TPA and saturation) except that the input Raman pump intensity is lowered to 48.5 MW/cm². Reduced Raman pump requirements and an increased logarithm dynamic range of 17.5 dB is gained at the expense of higher required signal intensity. The normalized-mean-square error is 0.8% and the maximum error is 3.4%.

At low input signal intensity, Raman pump source amplifies the output signal. The gain becomes less significant when the input signal grows, as the pump source is depleted by nonlinear absorption and amplification. Gain saturation modifies the input-output curve and expands the logarithmic region. Although it has a similar system setup, the logarithmic computing primitive functions fundamentally differently from a silicon Raman amplifier [11]. In the latter case, the signal intensity is significantly smaller than the pump. Under the assumption of negligible pump depletion, the output signal increases linearly with the input. In the logarithmic computing primitive case, both pump depletion and nonlinear absorption modify the output signal to be a sublinear function of the input.

As shown in Fig. 3, a 10.5 dB logarithmic region for signal input from 0.035 MW/cm² to 0.4 MW/cm² is achieved when the input Raman pump is 91 MW/cm². The introduction of the amplification significantly reduces the power requirement on the signal power, and also increases the logarithmic range.

A numerical sweep of the input pump intensity shows that at 48.5 MW/cm², the input logarithmic range is further expanded to 17.5 dB, from 0.4 MW/cm² to 22.4 MW/cm², as shown in Fig. 4. The Raman pump expands device flexibility, allowing one to trade higher Raman pump intensity for lower signal power and a larger logarithmic range.

Although use of a Raman pump can immensely reduce the required signal intensity (cf. Figs. 3 and 4), the output deviates from a logarithm at higher signal intensities. To shape the curve at high input intensity, a new pump source $I_P$ is injected into the waveguide to enhance the nonlinear absorption process through non-degenerate TPA with the signal wave. The evolution of signal wave $I_s$, Raman pump wave $I_R$, and non-degenerate TPA pump $I_P$ wave is described as:

$$\begin{cases} \dfrac{dI_s}{dz} = (-\alpha + g_R I_R)I_s - \beta_{TPA}(I_s + 2I_R + 2I_P)I_s - \sigma \Delta N I_s \\ \dfrac{dI_R}{dz} = \left(-\alpha - \dfrac{\lambda_s}{\lambda_R} g_R I_s\right) I_R - \beta_{TPA}(I_R + 2I_s)I_R - \sigma \Delta N I_R \\ \dfrac{dI_P}{dz} = -\alpha I_P - \beta_{TPA}(I_P + 2I_s)I_P - \sigma \Delta N I_P \\ \Delta N = \dfrac{\tau_c \beta_{TPA}}{2h\nu_0}(I_s^2 + I_R^2 + I_P^2 + 2I_s I_R + 2I_s I_P) \end{cases} \quad (6)$$

Note that no non-degenerate TPA between $I_R$ and $I_P$ is included in Eq. 6, which may be realized by simply the non-degenerate TPA pump wavelength

The non-degenerate TPA with the third beam $I_P$ suppresses the output signal as the input increases, extending the logarithmic range at the high-input side. For input Raman pump $I_R$ at 48.5 MW/cm² and the input non-degenerate TPA pump source $I_P$ at 55 MW/cm², the logarithmic input range is enlarged to 19.5 dB, from 0.32 MW/cm² to 28.8 MW/cm², as shown in Fig. 5.

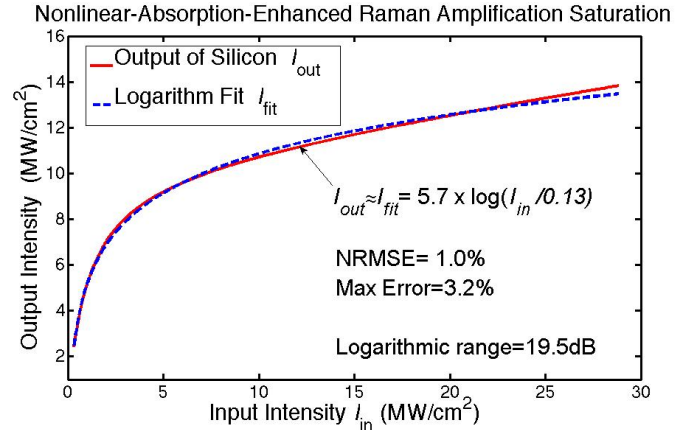

Fig. 5. Synthesis of the logarithmic computing primitive with the nonlinear-absorption-enhanced Raman amplification. The input Raman pump is 48.5 MW/cm² and the input non-degenerate TPA pump source is 55 MW/cm². The output is fit to a logarithmic function over a 19.5 dB input range with a normalized-mean-square error is 1.0% and the maximum error is 3.2%.

Compared to the synthesis of the logarithmic computing primitive, the scaling and the natural exponential computing primitive are implemented in a more straightforward way.

One way to perform the scaling function is to use a variable optical attenuator. A p-i-n diode structure fabricated on Si waveguide attenuates the input optical beams as a function of the current injection density [12,13].

Another way to implement the scaling primitive is to use the third-order parametric process. The output wave at a new frequency wave $\omega_4$ is generated from the mixing of the input waves at $\omega_1, \omega_2$, and $\omega_3$ as [14]:

$$\frac{dE_4}{dz} = -\frac{\alpha}{2}\epsilon_4 + \frac{2in_2\omega_4}{c} E_1 E_2 E_3^* e^{-i\Delta kz} \quad (7.1)$$

$$\rightarrow \frac{dI_4}{dz} = -\alpha I_4 + \frac{2n_2\omega_4}{c^2\epsilon_0 n}\sqrt{I_1 I_2 I_3 I_4}\ sin\phi \qquad (7.2)$$

where $E$ is the electric field amplitude, $\Delta k$ is the phase mismatch, $sin\phi = 1$ for perfect phase match, $n$ is the refractive index and $n_2$ is the nonlinear-index coefficient, which is proportional to the real part of third-order susceptibility. Under the low depletion assumption and perfect phase matching, Eq. (7.2) calculates the output intensity of $I_4$ at distance $l$ to scale with the input signal $I_1(0), I_2(0), I_3(0)$:

$$I_4(l) = \frac{4n_2^2\omega_4^2}{c^4\epsilon_0^2 n^2}\frac{\left(1 - exp\left(-\frac{\alpha l}{2}\right)\right)^2}{\alpha^2} I_1(0)I_2(0)I_3(0) \qquad (7.3)$$

The natural exponential computing primitive can be realized with the Raman amplification process. When the signal is significantly smaller than the pump source and the nonlinear absorption is negligible, the output signal is solved as [15]:

$$I_s(l) = I_s(0)\ exp(-\alpha l)\ exp\left(g_R L_{eff} I_R(0)\right) \qquad (8)$$

where $l$ is the length of the amplifier and $L_{eff} = (1 - exp(-\alpha l))/\alpha$ is the effective length.

Exploiting the nonlinear optical properties native to silicon, we show an approach to creating a logarithmic-analog co-processor in silicon photonics. By engineering the relative strength of Raman amplification and nonlinear absorption, the sublinear relationship between signal input and output is tuned to emulate a logarithmic function. The logarithmic computing primitive, together with a scaling primitive and a natural exponential primitive, can be used sequentially to realize the extremely nontrivial analog optical exponentiation operation.

This work was supported by the Office of Naval Research (ONR) MURI Program on Optical Computing. This work was performed under the auspices of the U.S. Department of Energy by Lawrence Livermore National Laboratory under contract DE-AC52-07NA27344.